\documentclass[review]{elsarticle}

\usepackage{lineno,hyperref}
\usepackage{booktabs}
\usepackage{multirow}
\usepackage{subfig}
\usepackage{amssymb}
\modulolinenumbers[5]

\journal{Journal of \LaTeX\ Templates}

\begin{document}

\begin{frontmatter}

\title{New data for the definition of neutron beams for Boron Neutron Capture Therapy}

\author[mymainaddress,mysecondaryaddress,mythirdaddress]{M. Mac\'ias\corref{mycorrespondingauthor}}
\cortext[mycorrespondingauthor]{Corresponding author}
\ead{mmacias4@us.es}
\author[mymainaddress,mysecondaryaddress]{B. Fern\'andez}
\author[mythirdaddress]{J. Praena}

\address[mymainaddress]{Centro Nacional de Aceleradores (Universidad de Sevilla, CSIC, Junta de Andaluc\'ia), Sevilla, Spain}
\address[mysecondaryaddress]{Departamento de F\'isica At\'omica, Molecular y Nuclear, Universidad de Sevilla, Sevilla, Spain}
\address[mythirdaddress]{Departamento de F\'isica At\'omica, Molecular y Nuclear, Universidad de Granada, Granada, Spain}

\begin{abstract}
Boron Neutron Capture Therapy (BNCT) is a neutron radiotherapy used to treat tumours cells previously doping with Boron-10. This therapy requires an epithermal neutron beam for the treatment of deep tumours and a thermal beam for shallow ones. Thanks to recent high-current commercial accelerators, Accelerator-Based Neutron Sources (ABNS) are competitive option for providing therapeutic neutron beams in hospitals. In this work, the neutron field generated by the $^7$Li(p,n)$^7$Be reaction at $1950~$keV is studied as neutron source in ABNS, being measured by the Time-Of-Flight (TOF) technique at HiSPANoS facility (Spain). Moreover, two Beam Shaping Assemblies (BSA) for deep and shallow tumour treatment, which are specially designed for the $1950~$keV neutron field, are evaluated for BNCT via Monte Carlo simulations (MCNP). Results in agreement with the International Atomic Energy Agency (IAEA) figures of merit endorse the use of this neutron field for BNCT.
\end{abstract}


\end{frontmatter}


\section{Introduction}

Nowadays, neutrons play an important role in the total dose during radiotherapy treatments. A maximisation of their biological effects is carried out under the Boron Neutron Capture Therapy. BNCT is a two-component therapeutic modality currently considered for the treatment of several tumours resistant to chemo- and radio-therapeutic methods. This therapy is a biochemically targeted radiotherapy based on nuclear capture and fission reactions, that occur when a stable isotope $^{10}$B is irradiated with low energy thermal neutrons to yield high Linear Energy Transfer (LET) $\alpha$-particles and recoiling $^7$Li nuclei \cite{bnctref2}.

The therapy exploits the nuclear reaction $^{10}$B(n,$\alpha$)$^7$Li. The total cross-section is $3,837~$barn and its Q-value is $+2.79~$MeV. The reaction proceeds by two pathways. In the $6.3$\% of cases, the Q-value energy is shared between the $\alpha$-particle ($1.78~$MeV) and the lithium ion ($1.01~$MeV). In the other $93.7$\%, the reaction leaves the $^7$Li-ion in an excited state, from which it decays immediately to its ground state, emitting a $0.428~$MeV $\gamma$-ray, while the remaining energy is shared between the $\alpha$-particle ($1.47~$MeV) and the lithium-ion ($0.84~$MeV). The average ranges are $9~\mu$m and $5~\mu$m in tissue for $\alpha$-particles and lithium-ions, respectively. These dimensions are comparable with cell size.

BNCT needs a large number of thermal neutrons at the tumour site. Their neutron sources have been limited to specially modified nuclear reactors. Therefore, low-energy high-intensity particle accelerators have been developed for BNCT since the 1980s \cite{ab_ref5, ABNS-BNCT1,ABNS-BNCT2,ABNS-BNCT3}. Several countries are developing accelerators to produce epithermal neutron beams for BNCT. Accelerator-Based BNCT (AB-BNCT) is being viewed worldwide as the future modality to start the era of in-hospital facilities, and a new dedicated industry is already on the market. For instance, Neutron Therapeutics Inc.\ \cite{nt} has developed an accelerator-based, in-hospital neutron source, which is composed by a $2.6~$MeV electrostatic proton accelerator and a rotating, solid lithium target for generating neutrons. Sumitomo Heavy Industries, Ltd.\ sited in Japan \cite{shi} or RaySearch Laboratories \cite{ray} are other examples of the new BNCT industry. 

Lithium is a metal with two isotopes ($^7$Li $92.5$\%, $^6$Li $7.5$\%) and a melting point at $180.54~^\circ$C. It is used in its natural or in its isotopic form $^7$Li to produce large quantities of fast neutrons of relatively low energy. The $^7$Li(p,n)$^7$Be has a Q-value of $-1.64~$MeV and a threshold of $1.88~$MeV. The peculiarity of this reaction is its cross-section that increases rapidly up to $\sim 270~$mb within the first $50~$keV. Therefore, the neutron yield is very high for proton energies just over the threshold. The low melting point makes it difficult to employ this target with high proton currents ($>~$mA), necessary for performing BNCT. However, a high-power ABNS based on a liquid-lithium target (LiLiT) and the $^7$Li(p,n) reaction was developed at SARAF (Soreq Applied Research Accelerator Facility, Israel) as a prototype for BNCT \cite{ab_ref5}, overcoming the temperature handicap. 

This (p,n) reactions produce fast neutrons, which must be thermalised to be captured by the $^{10}$B nucleus being useful to therapy. In the treatment of shallow tumours, the original fast neutrons have to be shifted to completely thermalise. In the case of deep tumours, Monte Carlo simulations endorse the optimal energy to reside in the epithermal region at the patient's entrance \cite{bnctnicola0}. Neutrons have to be shifted to lower energies to ensure that the tissue between the skin and the tumour can completely thermalise them. The energy shifter, called Beam Shaping Assembly (BSA), is a stack of different materials, appropriately shaped to thermalise the original fast neutron field into the appropriate thermal or epithermal neutron beam. Therapeutic neutron beams with high spectral purity in these energy ranges could be produced with ABNS through a suitable neutron-producing reaction. A study by Bisceglie et al.\ \cite{bnctnicola} on different neutron producing reactions that could be used in conjunction with high-current accelerators to produce epithermal neutron beams for BNCT showed as a good candidate the epithermal neutron spectrum that can be produced with the $^7$Li(p,n) reaction at $1950~$keV with an appropriated neutron beam shaping.

To minimise undesirable radiation, which releases energy in healthy tissues upstream and downstream the tumour site, the International Atomic Energy Agency (IAEA) proposes figures of merit for shallow and deep tumours treatments. Given the following three energetic groups for neutrons: thermal (E$_n<0.5~$eV), epithermal ($0.5~$eV$~<~$E$_n <10~$keV), and fast (E$_n > 10~$keV), IAEA recommends the following neutron fluence ($\Phi$) rates and neutron absorbed dose (\.{D}) rates for shallow tumours: $\Phi_{th}~$ $\geqslant 10^9$cm$^{-2}$s$^{-1}$, $\Phi_{th}/\Phi_{tot}$ $\geqslant 0.9$, and \.{D}$_{epi+fast}/\Phi_{th}~$ $\leqslant 2 \times 10^{-13}$Gy\,cm$^2$. For deep tumours: $\Phi_{th}~$ $\geqslant 10^9$cm$^{-2}$s$^{-1}$, $\Phi_{epi}/\Phi_{th}$ $\geqslant 100$, $\Phi_{epi}/\Phi_{fast}$ $\geqslant 20$, and \.{D}$_{fast}/\Phi_{epi}~$ $\leqslant 2 \times 10^{-13}$Gy\,cm$^2$.

In this work, we carried out a deep study of the epithermal neutron beam production by $^7$Li(p,n) reaction at $1950~$keV with the intent to check its viability as neutron source for BNCT. Measurement and data analysis of the mentioned neutron field are presented in Sections \ref{exp} and \ref{ana} respectively. Our results are presented in Section \ref{resu} compared to an analytical descriptions of the  reaction proposed by Lee \& Zhou \cite{Lee} (based on Liskien \& Paulsen nuclear data \cite{LiskienP}). Implications for the development of therapeutic neutron beams for BNCT from a neutron field are discussed in Section \ref{imp}.

\section{Experimental setup and measurement}
\label{exp}
The experiment was performed at HiSPAlis Neutron Source (HiSPANoS) facility, at National Accelerator Centre (CNA) \cite{CNA}. HiSPANoS facility is the first ABNS in Spain. This facility provides different neutron fields in continuous-wave employing $^7$Li(p,n)$^7$Be and $^2$H(d,n)$^3$He reactions. Since 2019, a new pulsed mode and a new experimental Time-Of-Flight (TOF) line are available \cite{macias0,MACIAS}, providing proton and deuteron pulses of $1~$ns for neutron generation. 

\begin{figure}[ht]
\centering
\includegraphics[width=.5\textwidth]{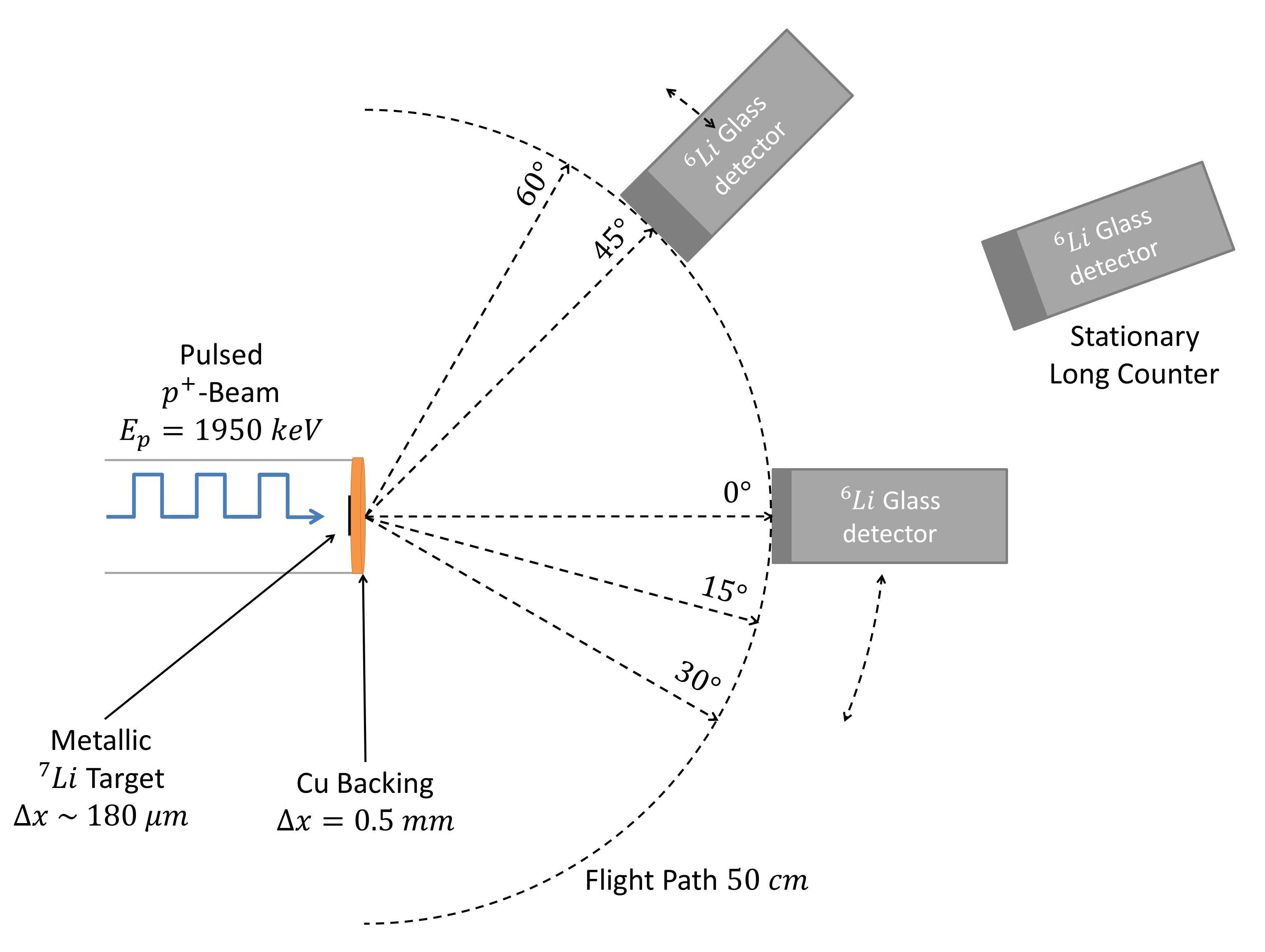}
\caption{Scheme of the experimental setup. The target assembly, target, and detectors distribution are shown.}
\label{fig:setup}
\end{figure}

For the neutron production, a proton pulsed beam of $1950\pm 1~$keV with $62.5~$kHz of repetition rate, $1~$ns pulse width, and a stable average current of $12~$nA was employed. A target assembly specially designed to minimise the neutron scattering in the target surroundings was used. It consists of a circular copper backing ($4~$cm diameter and $0.5~$mm thickness), an aluminium cylinder ($4.2~$cm diameter and $15~$cm long), and an O-ring for their joining. A thick lithium-metallic target was prepared by forced pressure onto the backing in inner gas. The sample had a diameter of $1~$cm and a thickness of about $180~\mu$m, therefore protons were completely stopped. In order to protect the Li sample from oxidation, the target was mounted on the beamline keeping an Argon atmosphere. After its installation, the Argon was extracted by the vacuum system until a stable work pressure of $10^{-6}~$mbar. To avoid the lithium target melting due to the impact of proton pulses, a jet of air cooled the target. The detection system consisted of three $^6$Li-glass detectors and an aluminium goniometer for a precise location in the angle of the detectors respect to the target completed the experimental setup. The target was located at the center of the goniometer with movable stands for placing the detectors. A movable detector ($5.08~$cm diameter and $2.54~$cm thick) recorded the signals at $45^\circ$ and $60^\circ$. A second movable detector (5.08-cm diameter and 1.27-cm thick) recorded the signals at $0^\circ$, $15^\circ$, and $30^\circ$. Both detectors were placed at $50~$cm (flight path), at the same target height. A third $^6$Li-glass detector was used as a stationary long counter in a fixed position without geometrical interference with the others. Figure \ref{fig:setup} shows a scheme of the experimental setup.

The detector signals were processed with a CAEN\textregistered\ data acquisition (DAQ) model DT5730 with 14 bit at 500 Ms/s FLASH ADC Waveform Digitizer \cite{dt}. The proton pulsed beam induces a signal per pulse in a pick-up installed at the experimental TOF line. These were used as the start signals for the acquisition. Then, the DAQ was programmed to open a time acquisition window of $1500~$ns, collecting the signals from the detectors. The trigger enables the event building, that includes the waveforms (i.e. the raw samples), the trigger time, the baseline, and the Integral Charge (IC) of the signals \cite{coinci}. Data are continuously written in a circular memory buffer. When the trigger occurs, the digitizer writes further samples for the post-trigger and freezes the buffer that can be read by one of the provided readout links. This acquisition can be acquired continuously without dead time in a new buffer \cite{dt}.

\begin{figure}[ht]
\centering
\includegraphics[width=.8\textwidth]{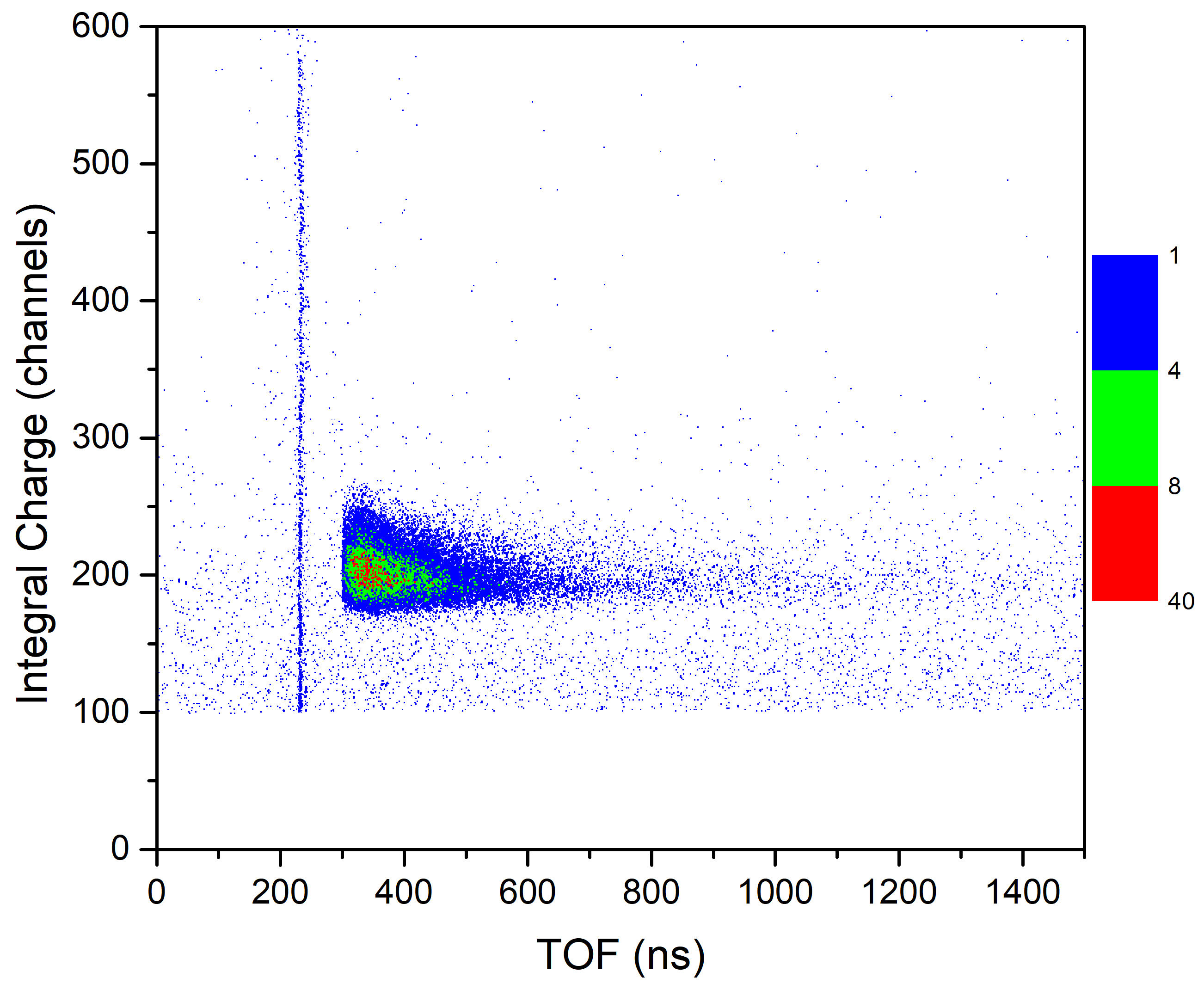}
\caption{2D histogram where IC vs.\ TOF is plotted. The $\gamma$-rays appear as a vertical line and the neutrons appear in the horizontal drop. The low frequency allowed the neutrons scattering to disappear over the detector between pulses, as can be seen before the neutron drop.}
\label{fig:2D_1950}
\end{figure}

The raw data were accumulated in a two-dimensional histogram consisting of TOF versus IC, as shown in Figure \ref{fig:2D_1950}. The $\gamma$-rays appear as a vertical line in time close to $240~$ns. The neutrons appear in the horizontal band from $320~$ns with a drop shape. It can be noticed the excellent discrimination between $\gamma$-rays and neutrons during the experiment. The neutron scattering component of the background is identified between the $\gamma$-flash and the onset of the neutron drop. The low repetition rate allowed minimising the neutrons scattering overlap of successive neutron bursts, as shown in the area before the $\gamma$-flash. Finally, it can be noticed that background signals are scattered overall area with an apparent time-independence.

\begin{figure}[ht]
\centering
\subfloat[$0^\circ$]{\includegraphics[width=.5\textwidth]{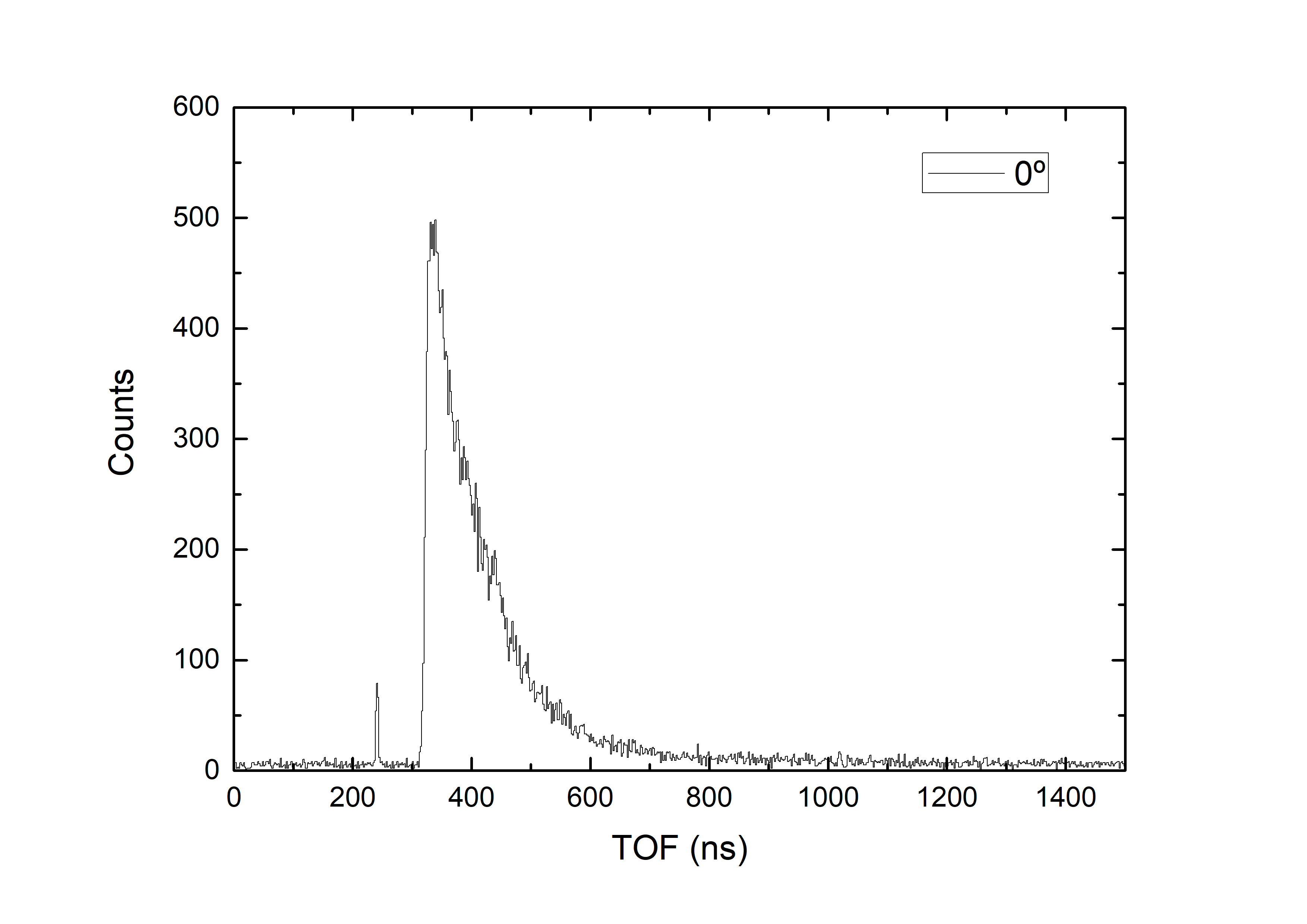}}
\subfloat[$15^\circ$]{\includegraphics[width=.5\textwidth]{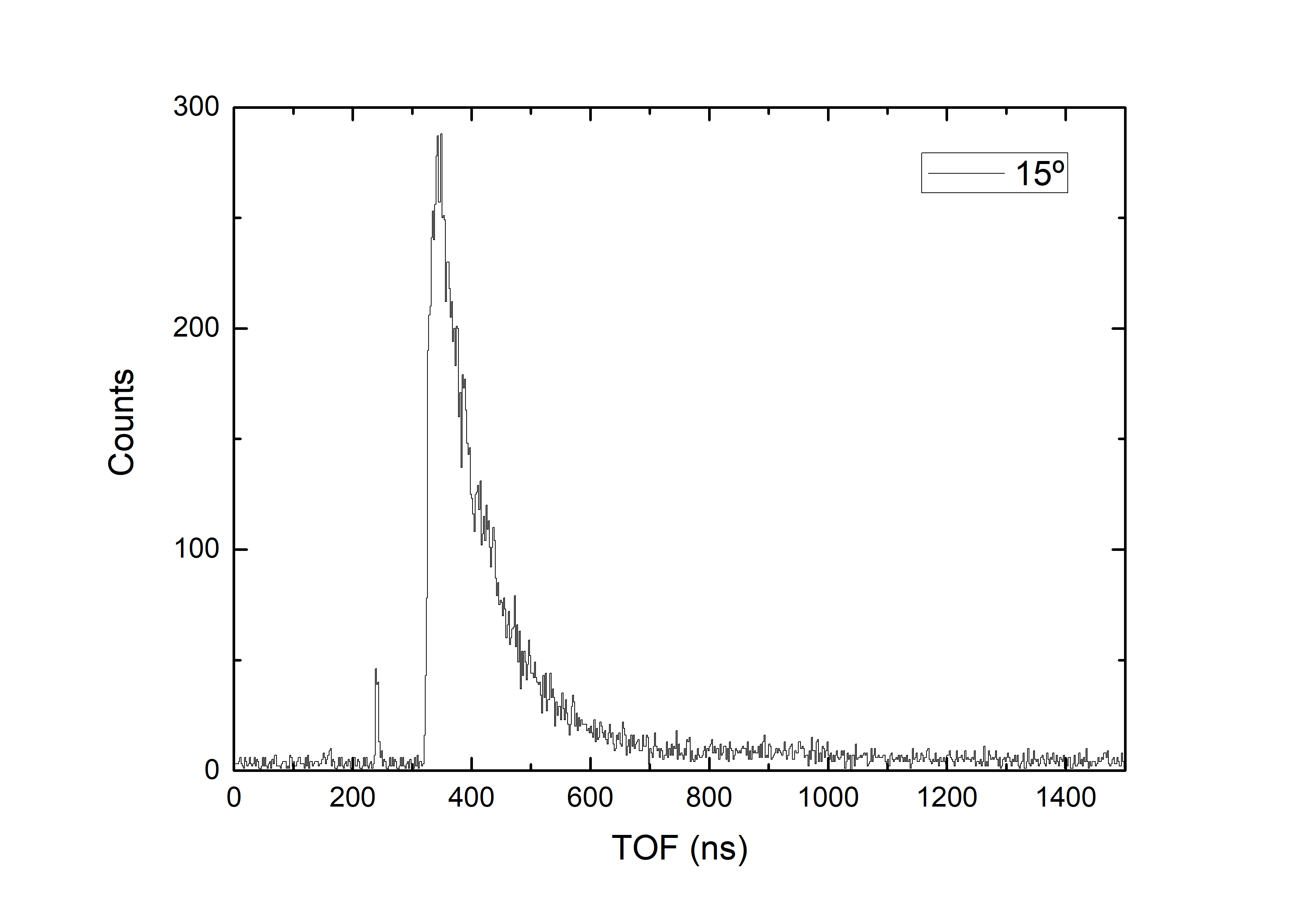}}
\vspace{0.1cm}
\subfloat[$30^\circ$]{\includegraphics[width=.5\textwidth]{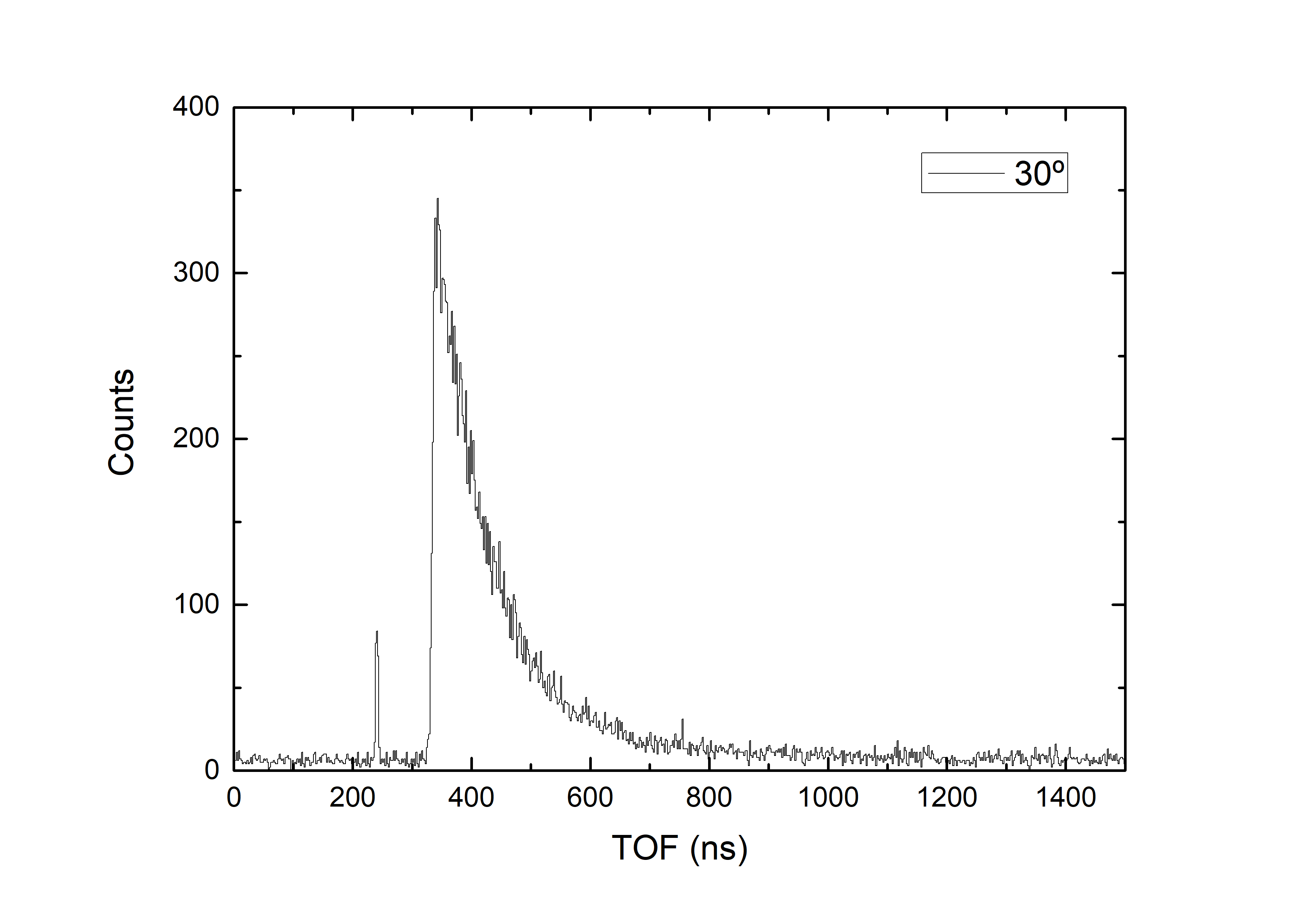}}
\subfloat[$45^\circ$]{\includegraphics[width=.5\textwidth]{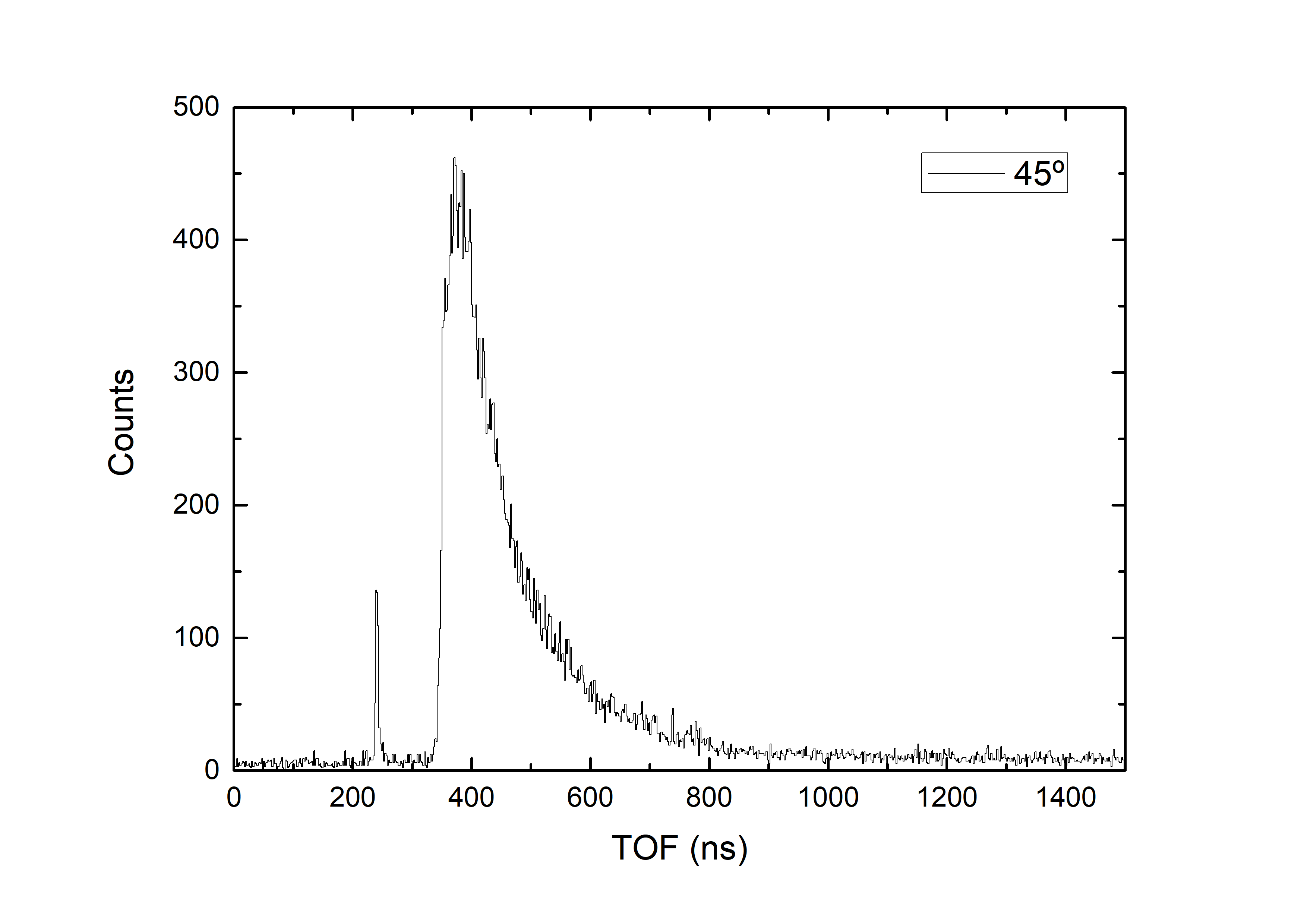}}
\vspace{0.1cm}
\subfloat[$60^\circ$]{\includegraphics[width=.5\textwidth]{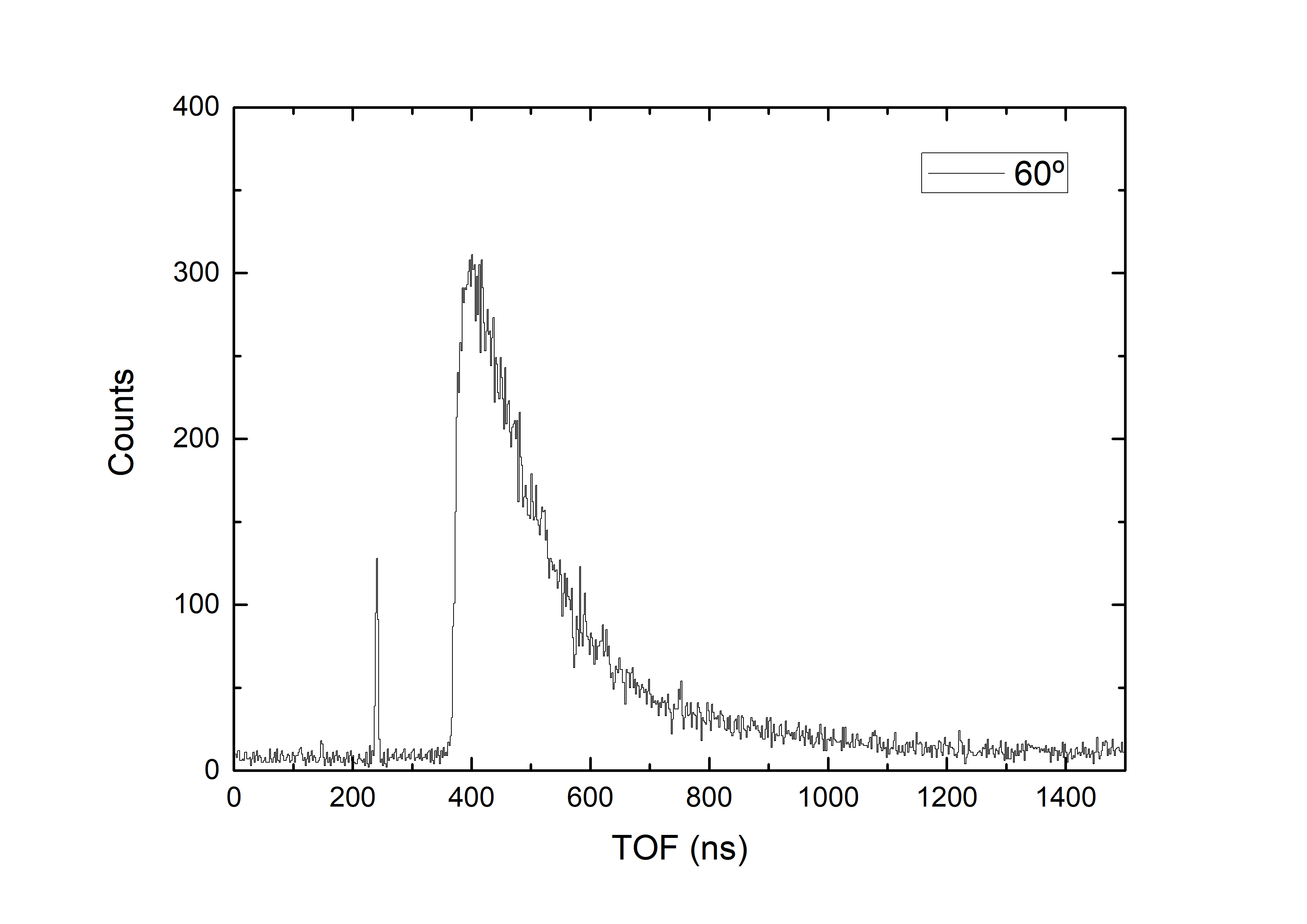}}
\subfloat[$0^\circ$ $+$ shadow bar]{\includegraphics[width=.5\textwidth]{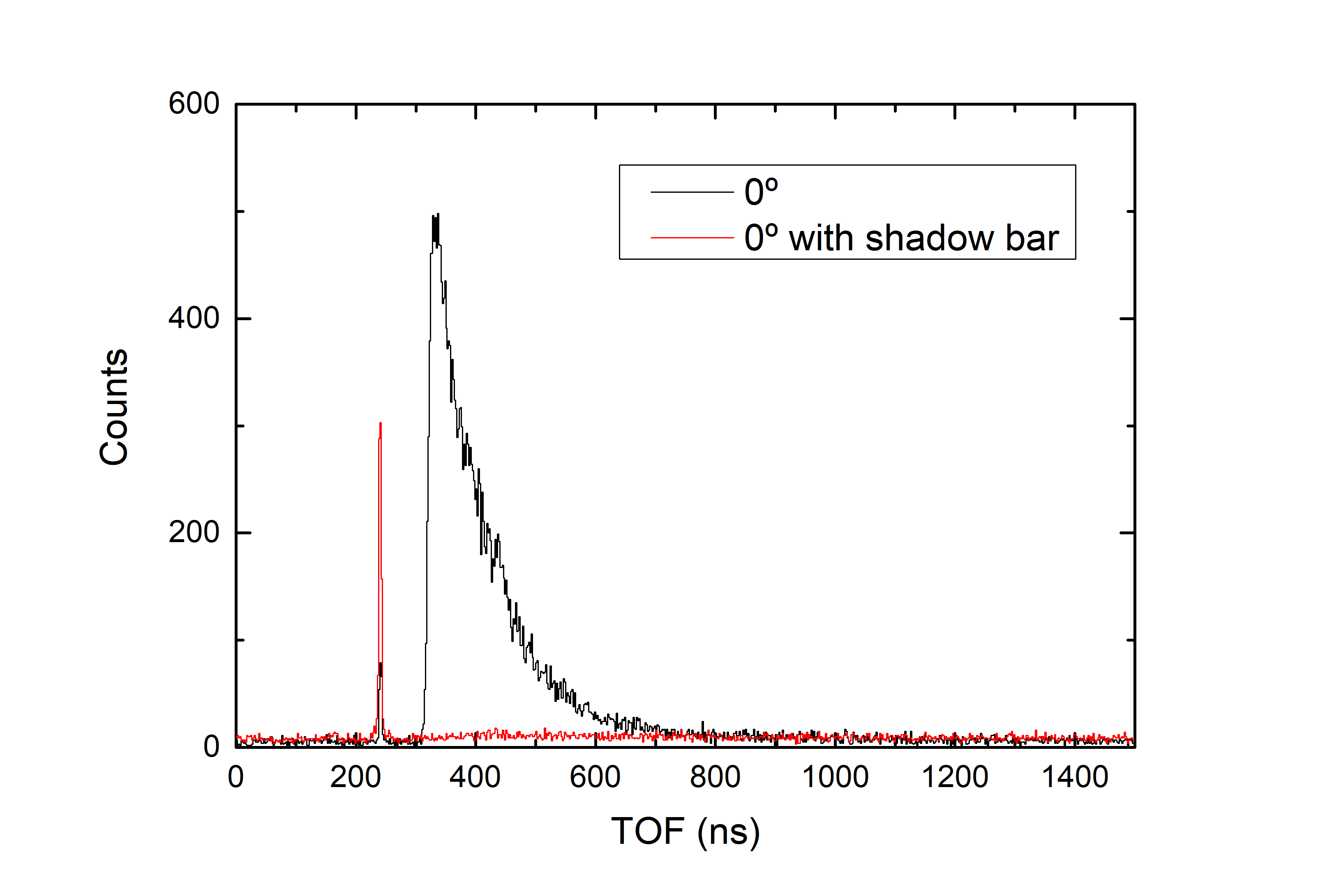}}
\caption{TOF spectra of raw data at $62.5~$kHz with a $50~$cm flight path at $0^\circ$, $15^\circ$, $30^\circ$, $45^\circ$, $60^\circ$, and a comparison between $0^\circ$ TOF and $0^\circ$ TOF where the detector is covered with the shadow bar.}
\label{fig:tofs_1950}
\end{figure}

The quality of the TOF measurements is illustrated in Figure \ref{fig:tofs_1950}, which shows the spectrum taken at different angles and $0^\circ$ with shadow bar. The time resolution in TOF is determined by the Gaussian fits of each $\gamma$-peak, which includes the time spread of the proton beam pulses, the time resolution of the $^6$Li-glass detector, and the time resolution of the DAQ. A common $\sigma$ value of $2~$ns was obtained, thus the total time resolution of the experiment.

\begin{figure}[ht]
\centering
\includegraphics[width=.6\textwidth]{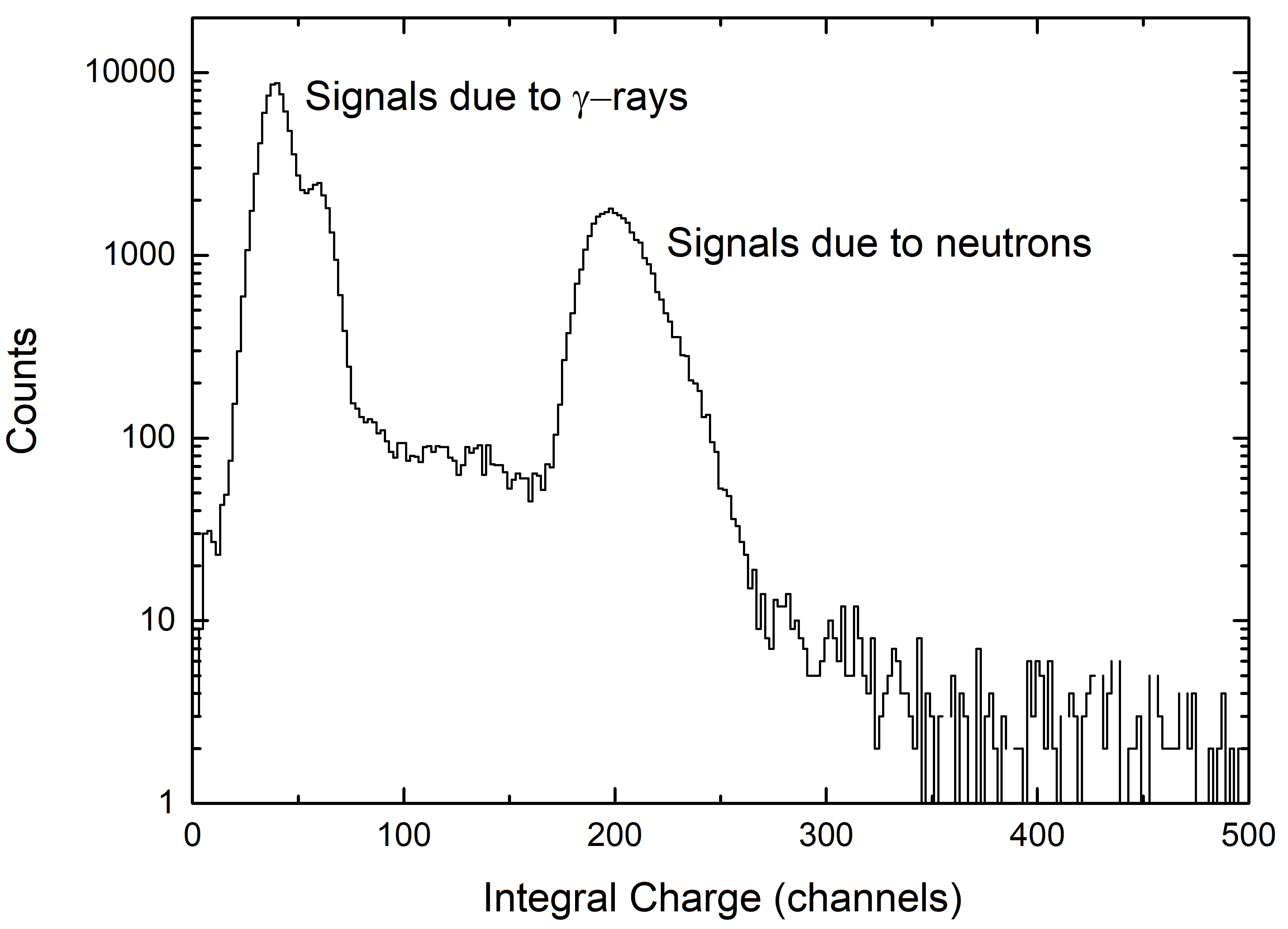}
\caption{Raw data organized in a one-dimensional IC histogram. Signals corresponding to $\gamma$-rays and neutrons are indicated.}
\label{fig:identif_Qlong_50}
\end{figure}

In order to extract the background and determine its possible correlation with the TOF, the detectors were shielded with shadow bars and the TOF spectrum was recorded at each angular position. The shadow bars were a polyethylene cylinder with $5.08~$cm diameter and $20~$cm length. In Figure \ref{fig:tofs_1950}(f), a comparison between TOF histograms at $0^\circ$ without shadow bar (black line) and with shadow bar (red line) is shown. The signals with shadow bar (red line) were due to the prompt $\gamma$-rays produced in the $^7$Li(p,n)$^7$Be directly detected ($\gamma$-flash) and scattered neutrons and $\gamma$-rays in the experimental hall and the environment background. As can be seen, the background was uncorrelated with the TOF (flat background) and very low. The background level was also reduced by setting an IC threshold (channel $100$) containing mostly low-energy $\gamma$-events and electronic noise as shown in the IC histogram of Figure \ref{fig:identif_Qlong_50}.

\section{Data Analysis}
\label{ana}
The neutron detection pathway of $^6$Li-glass scintillators is defined by the $^6$Li(n,t)$^4$He reaction. However, the presence of resonances in some structural materials of the detector in the measured energy range, as $^{56}$Fe and $^{28}$Si, provided small contributions to the neutron efficiency. These possible contributions was studied via the resolution function of the experiment using a detailed model of the detectors purchased to Scionix Ltd.\ \cite{liglass} and Monte Carlo simulations with MCNPX \cite{mcnpx}.

\begin{figure}[ht]
\centering
\includegraphics[width=.6\textwidth]{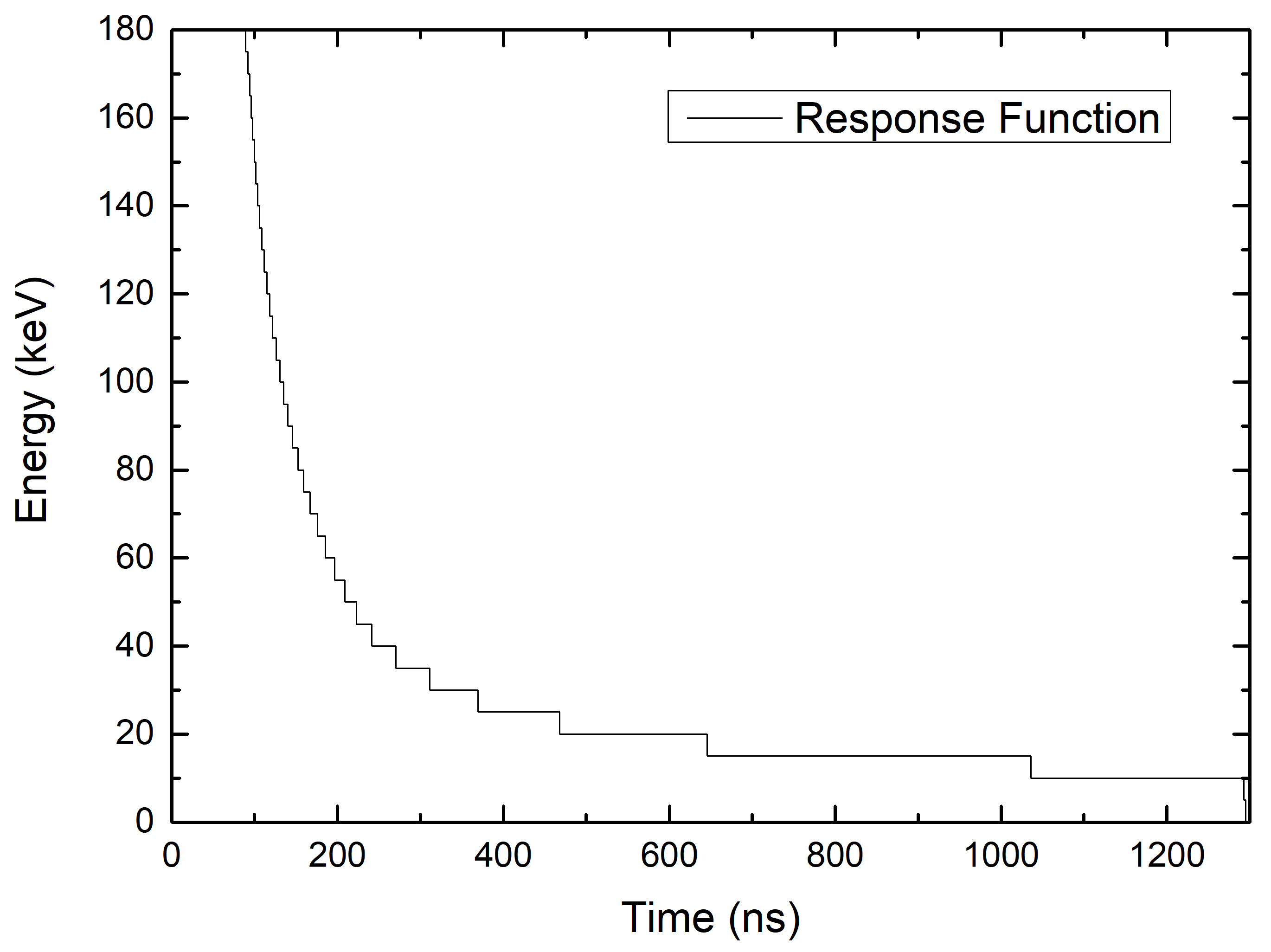}
\caption{The response function of the set-up determined the relationship between the TOF bins and each energy bin obtained.}
\label{fig:my_t_E_RF_1950}
\end{figure}

The statistical uncertainty dominated in our experiment as a consequence of the low repetition rate. A linear energy binning of $5~$keV steps was employed, being a good compromise between the statistical uncertainties and the energy resolution. 
The resolution function also takes into account all the possible interactions suffered by neutrons from the source until they induce $^6$Li(n,t)$^4$He reactions in the detector. These previous interactions, which generated an extra tof, are weighted and corrected by the response function on the TOF-to-Energy conversion. Figure \ref{fig:my_t_E_RF_1950} shows the resolution function associated with this experiment for the TOF-to-Energy conversion, where the detector efficiency, target, target assembly, setup distribution, and neutron energy range are included.

\begin{figure}[ht]
\centering
\includegraphics[width=.6\textwidth]{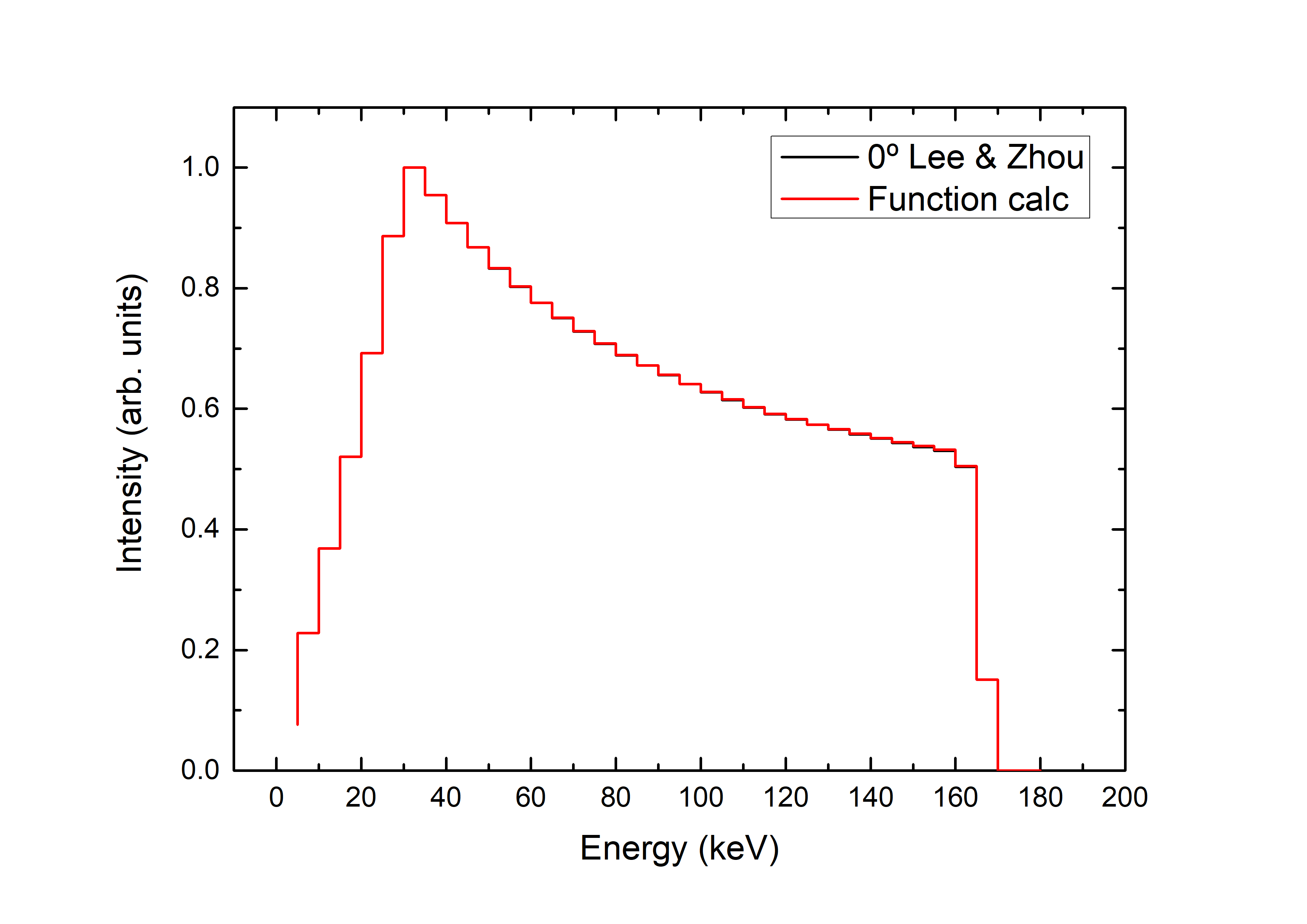}
\caption{Comparison between Lee \& Zhou field at $0^\circ$ \cite{Lee} and the response function conversion of a simulated TOF, both represented in the same binning and normalised by their sums.}
\label{fig:my_t_E_1950}
\end{figure}

The uncertainty related to the resolution function was estimated by the comparison of two energy distributions for each angle position and detector type. The first distribution was obtained from the simulation of the TOF spectrum scored by a detector and the application of the resolution function. For this, in the input file of the MCNPX code \cite{mcnpx}, the analytical descriptions of the $^7$Li(p,n)$^7$Be reaction at E$_p$=$1950~$keV of Lee and Zhou \cite{Lee} was introduced. This simulation recorded in a time histogram the invested time by the neutrons between the source and the $^6$Li(n,t)$^4$He reaction point inside the detector. The second energy distribution was obtained from the simulation of the energy distribution directly recorded by the detector using the same analytical descriptions of the neutron field. Figure \ref{fig:my_t_E_1950} shows the comparison between the resulting energy distributions. The comparison showed a negligible impact on the total uncertainty; however a conservative 1\% was assumed. 

\section{Results}
\label{resu}

\begin{figure}[ht]
\centering
\centering
\subfloat[$0^\circ$]{\includegraphics[width=.5\textwidth]{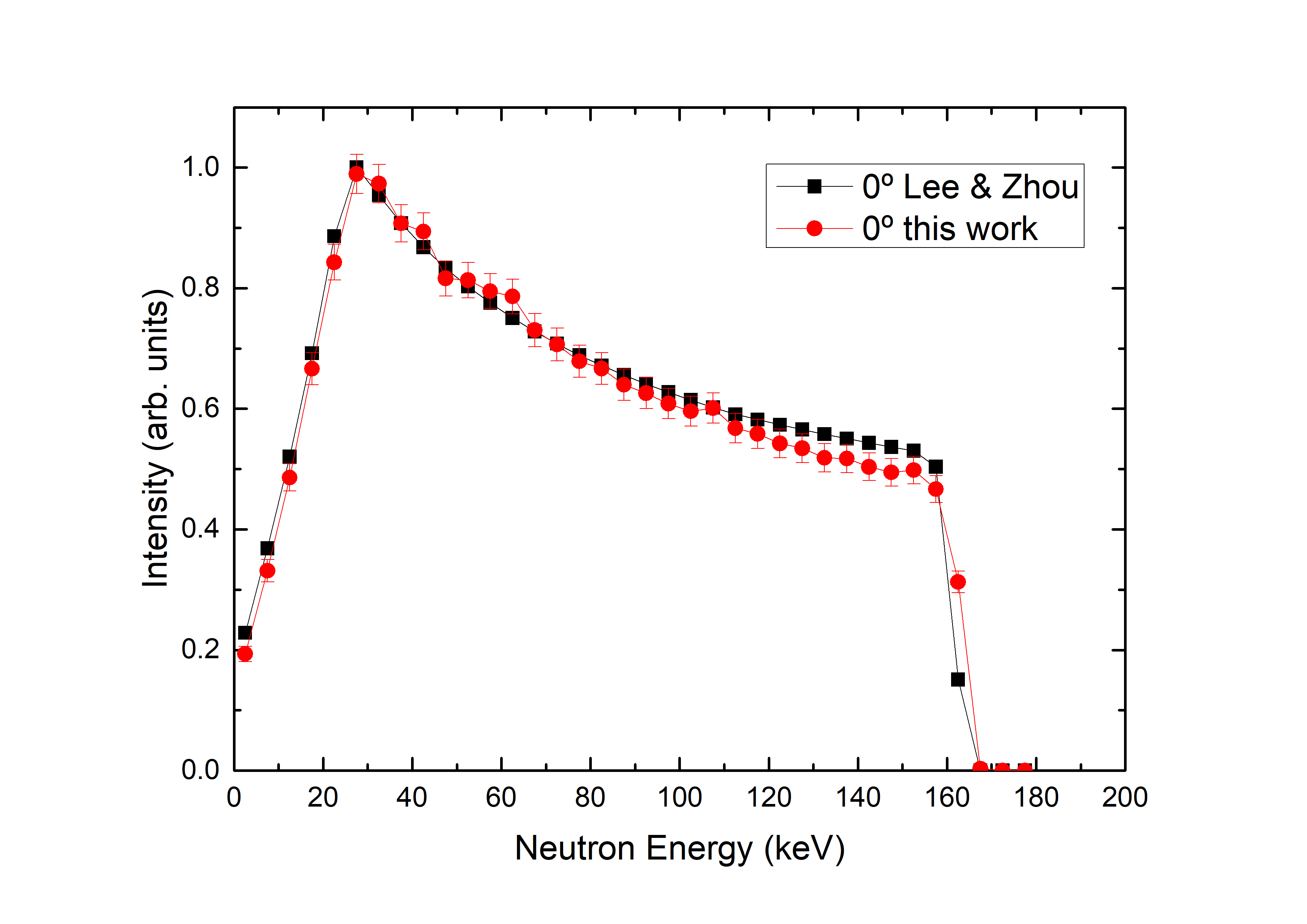}}
\subfloat[$15^\circ$]{\includegraphics[width=.5\textwidth]{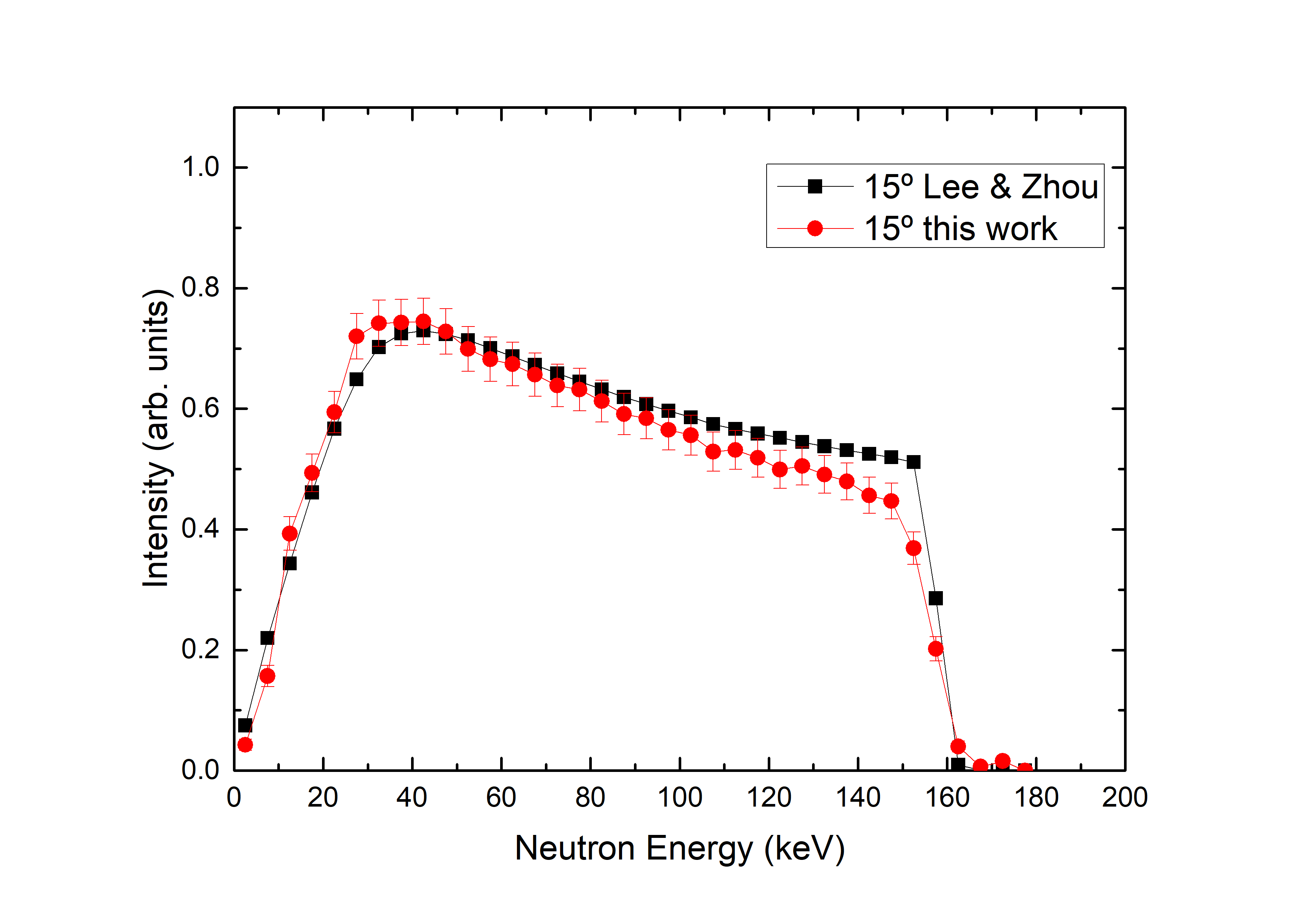}}
\vspace{0.1cm}
\subfloat[$30^\circ$]{\includegraphics[width=.5\textwidth]{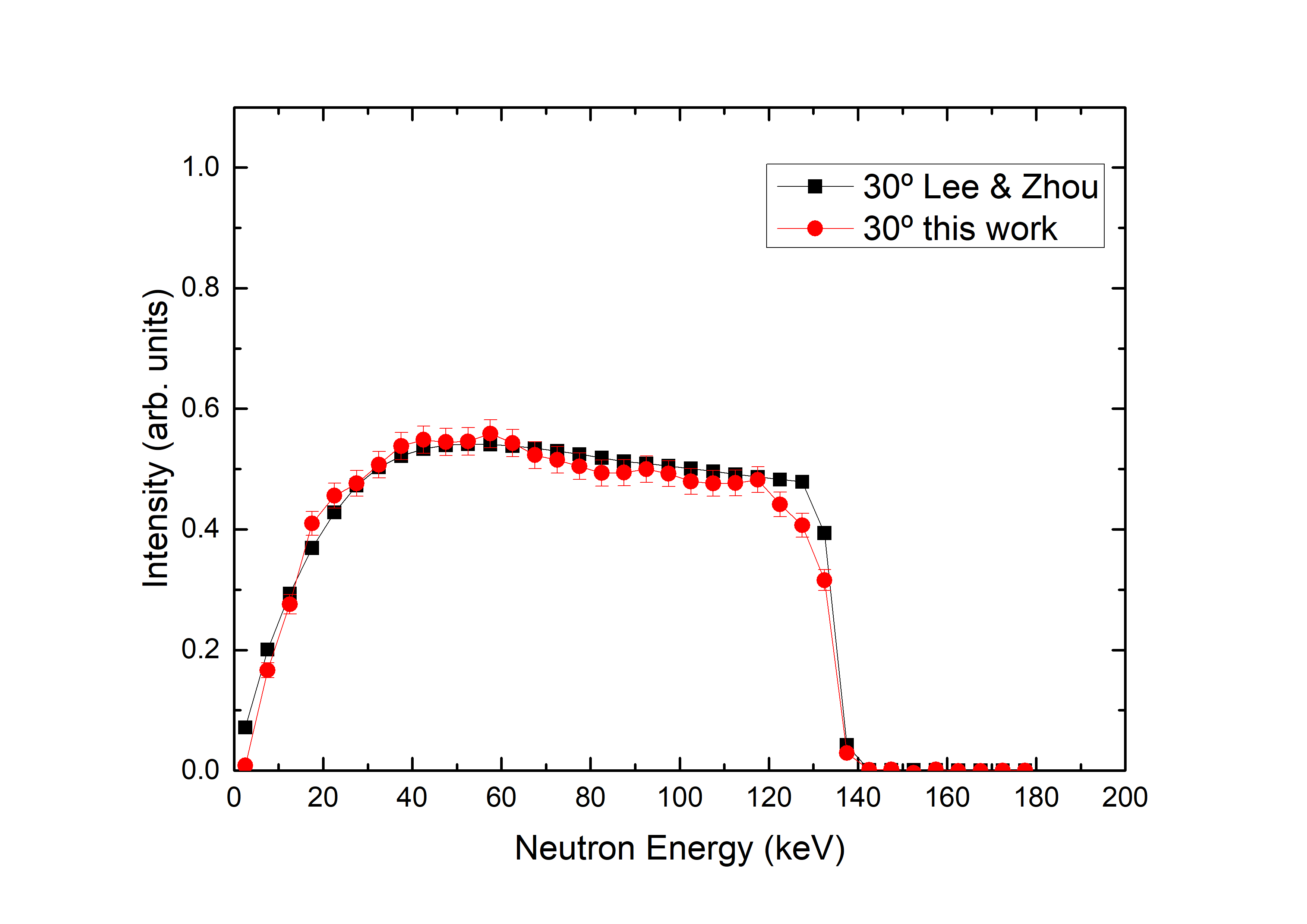}}
\subfloat[$45^\circ$]{\includegraphics[width=.5\textwidth]{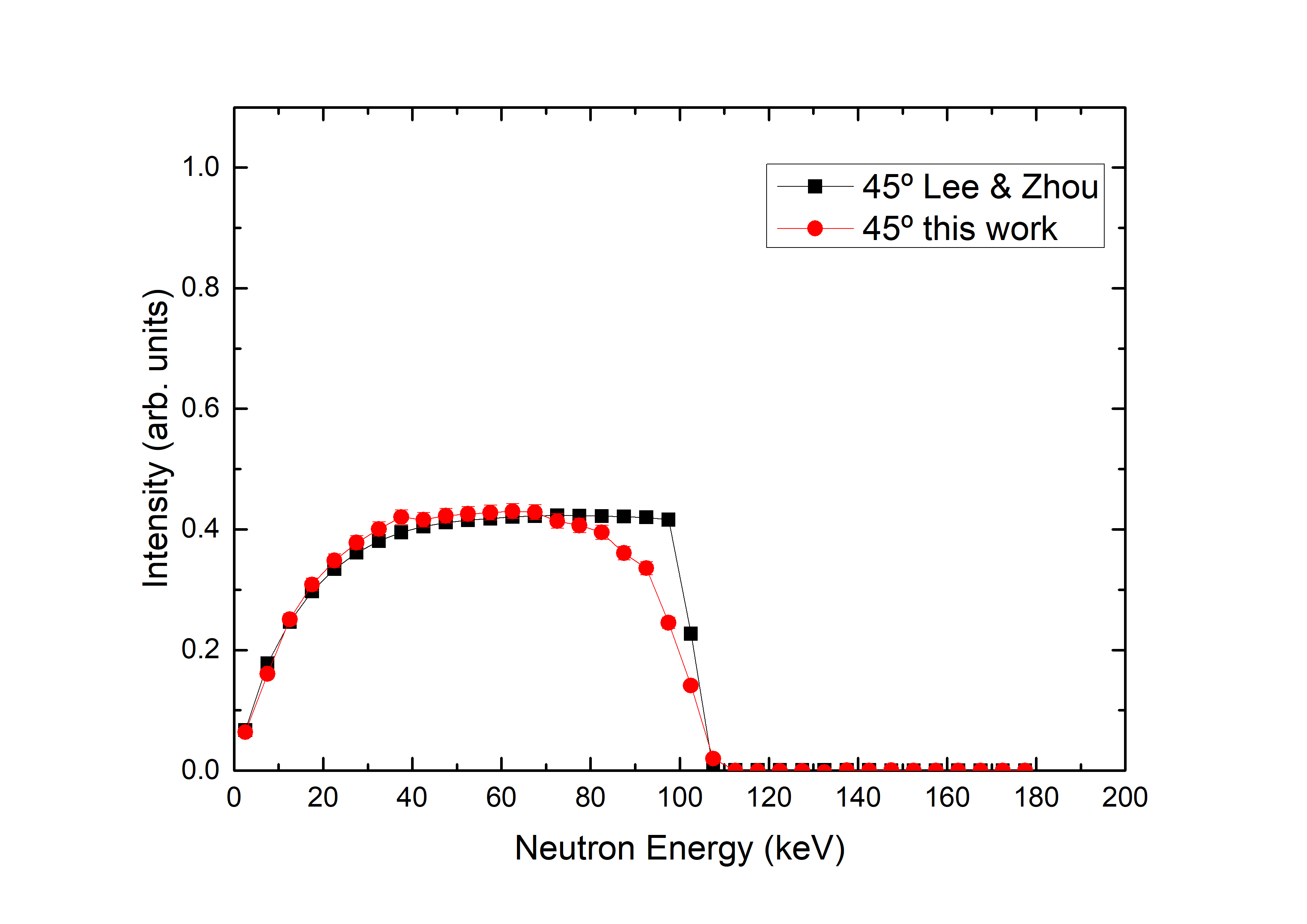}}
\vspace{0.1cm}
\subfloat[$60^\circ$]{\includegraphics[width=.5\textwidth]{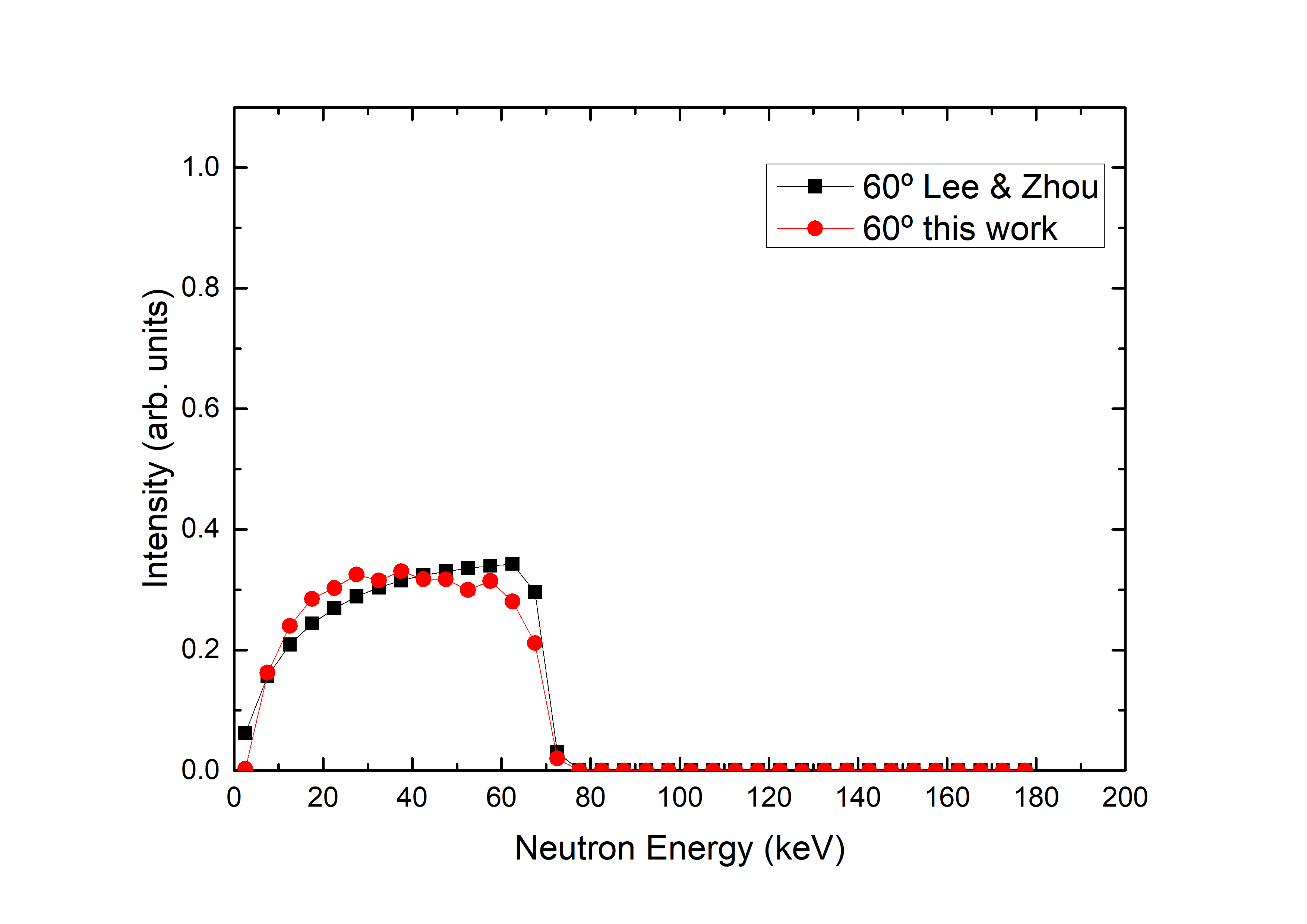}}
\caption{Experimental neutron spectra from $0^{\circ}$ to $60^{\circ}$ in steps of $15^{\circ}$. 0-deg spectrum is normalized by the maximum to improve the comparison between different angle spectra. Red lines correspond to this work and black lines correspond to Lee \& Zhou field \cite{Lee} in bins of $5~$keV. Uncertainties are statistical.}
\label{fig:intensity_1950}
\end{figure}

The spectrum of each measured emission angle are plotted and compared with their respectively analytical solution by Lee \& Zhou \cite{Lee} in Figure \ref{fig:intensity_1950}. Note that the comparison consists of two aspects: the spectra shape and their relative intensity. Also, particular attention should be paid to the maximum neutron energy of each spectrum. Regarding the intensity of the neutrons: it depends on the experimental set-up, so it is necessary to normalize our $0^{\circ}$ histogram and the Lee \& Zhou $0^{\circ}$ histogram by the maximum. The rest of our spectra are obtained directly from their relative intensity to our normalized $0^{\circ}$ spectrum, and then, they are compared with the spectra of Lee \& Zhou without further normalization.

Also, for a proper comparison of the neutron spectra from $0^{\circ}$ to $60^{\circ}$, each Lee \& Zhou energy spectrum has to be scored by the corresponding detector for each angle at $50~$cm in MCNPX simulations. As commented before, in order to generate the angle and energy distribution of neutrons at E$_p=1950~$keV the analytical description of Lee \& Zhou \cite{Lee} based on Liskien \& Paulsen nuclear data \cite{LiskienP} was used. Figure \ref{fig:intensity_1950} shows all the spectra of the present work normalised (red lines) and compared to Lee \& Zhou field (black lines). The agreement in the shape of the spectra and the maximum neutron energy is good within statistical uncertainties. The differences at high energies are justified by the comparison with an analytical description. Facts like time resolution of the acquisition system or detector efficiencies are not taken into account. Table \ref{tab:E_1950} shows a comparison between Lee \& Zhou spectra and our results of the integral neutron field per angle, relative to its respective angular-integral spectrum. The agreement in their relative intensity is excellent. The comparison of our results with Lee \& Zhou confirms the good performance of the experiment at HiSPANoS facility and represents the first nuclear data production (after the commissioning \cite{MACIAS}) with the TOF technique at the CNA.

\begin{table}[ht]
\centering
\footnotesize
 \begin{tabular}{ccc}
  \toprule
Angle & Lee \& Zhou & this work \\
  \midrule
$0^{\circ}$  & 33.6 & 33.9 \\
$15^{\circ}$ & 28.6 & 28.4 \\
$30^{\circ}$ & 19.9 & 20.0 \\
$45^{\circ}$ & 11.9 & 11.7 \\
$60^{\circ}$ &  6.1 &  6.1 \\
 \bottomrule
 \end{tabular}
 \caption{Integral neutron field per angle comparison between Lee \& Zhou spectra and our results, relative to its respective angular-integral spectrum.}
 \label{tab:E_1950} 
\end{table}


\section{Implications for BNCT}
\label{imp}
The advancement of BNCT requires suitable neutron sources for installation in hospital environments. AB-BNCT facilities are more appropriate in specialised health-care institutions than reactor-based neutron sources because of their lower prize, size, and radiation hazards; and larger simplicity in bureaucracy, installation, and maintenance. 

\begin{figure}[ht]
\centering
\subfloat[Shallow tumours]{\includegraphics[width=.5\textwidth]{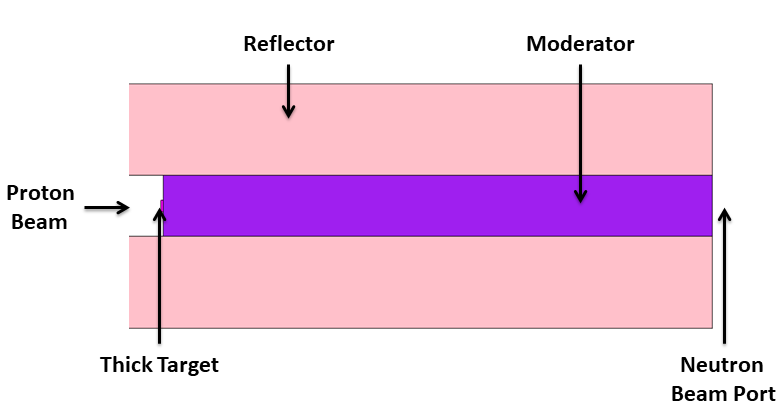}}
\subfloat[Deep tumours]{\includegraphics[width=.5\textwidth]{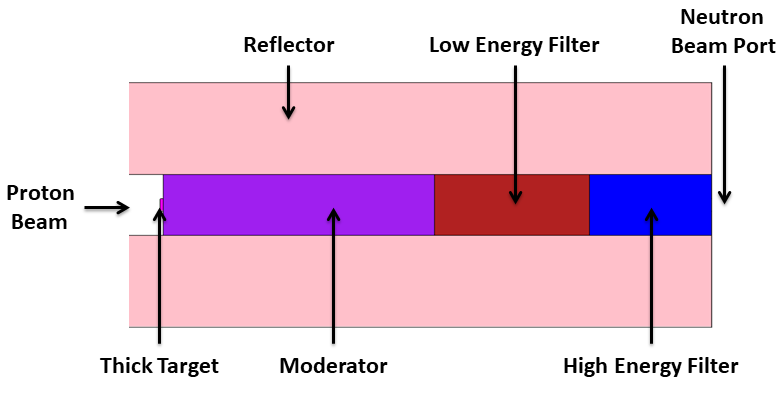}}
\caption{A schematic view of the final BSA configurations for shallow (a) and deep (b) tumours is shown. In (a), a combination of Pb as a reflector (pink) and Polyethylene as a moderator (violet) was the optimal configuration for the thermal neutron beam production. In (b), a combination of Polyethylene as a moderator (violet), Hexachlorobenzene (C$_6$Cl$_6$) as a low energy filter (red), Titanium (Ti) as a high energy filter (blue), and Pb as a reflector (pink) was the optimal configuration for the epithermal neutron beam production.}
\label{fig:bsa}
\end{figure}

In this work, we have developed two BSAs for the treatment of shallow and deep tumours using the previously measured neutron field. A thick lithium target was considered in designing the BSA configurations shown in Figures \ref{fig:bsa} (a) and (b).

\begin{figure}[ht]
\centering
\subfloat[Shallow tumours]{\includegraphics[width=.5\textwidth]{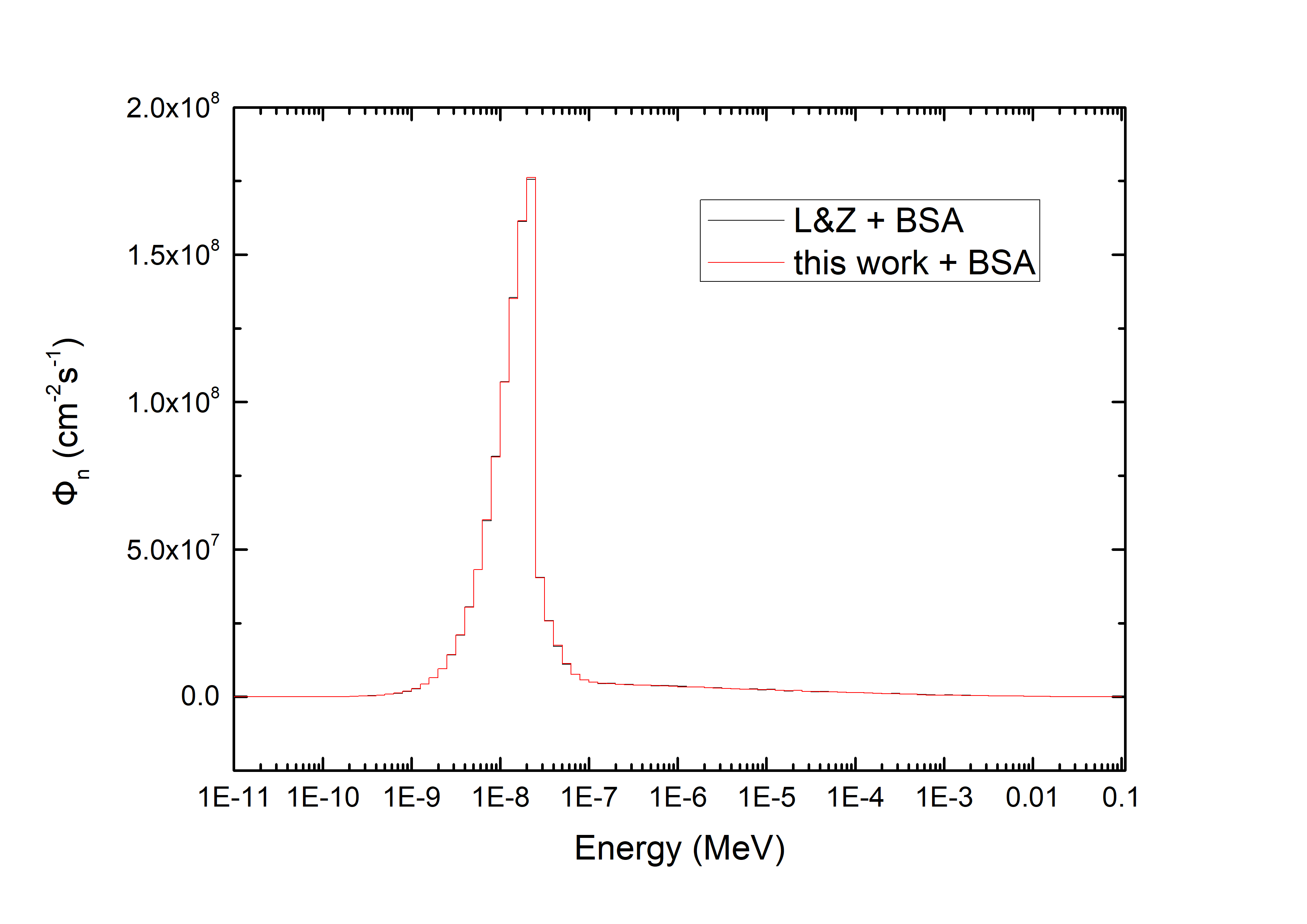}}
\subfloat[Deep tumours]{\includegraphics[width=.5\textwidth]{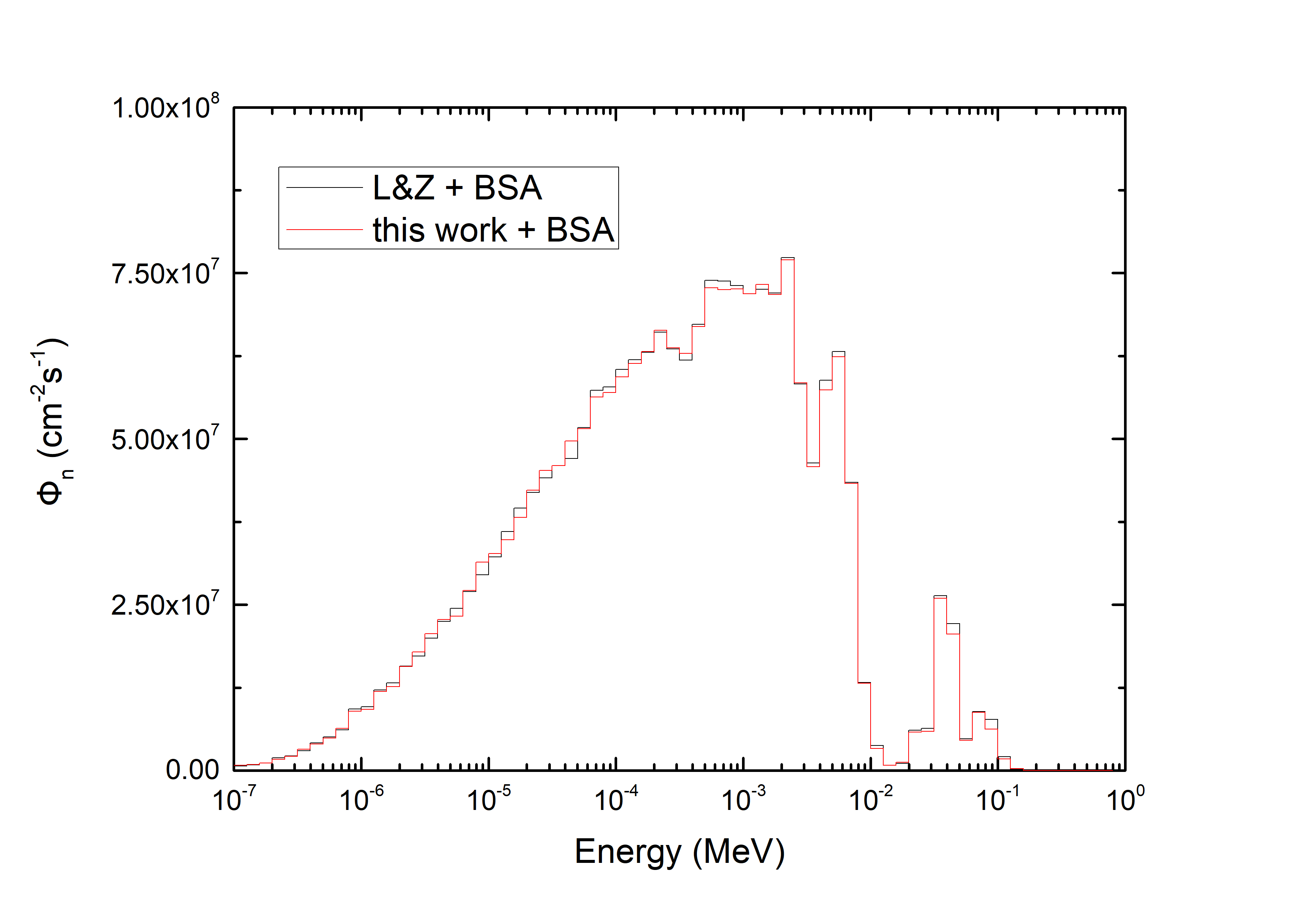}}
\caption{The neutron flux-energy spectra at $1~$cm from the BSA beam ports for shallow (a) and deep (b) tumours with $81~$mA and $85~$mA proton currents respectively.}
\label{fig:espectro}
\end{figure}

In the shallow BSA case, different polymers were tested as moderators, being Polyethylene the better option to scatter neutrons to lower energies. Lead (Pb) was employed as neutron reflector material. The Monte Carlo code MCNPX \cite{mcnpx} was used to perform the design calculations. A combination of Pb as a reflector and Polyethylene as a moderator was the optimal configuration for the treatment with $1950~$keV of initial proton energy. The main dimensions of the BSA are shown in Table \ref{tab:bsa_design}. The tumour skin was positioned at $1~$cm to the BSA. With a proton current of $81~$mA over a Lithium-metallic sample, this work is in agreement with the IAEA neutron figures of merit recommendations for shallow tumours. The analytical description and our experimental results and IAEA figures of merit are displayed in Table \ref{tab:bsa_IAEA}. The good agreement is also visible in the comparison of IAEA neutron flux-energy spectrum in the tumour area shown in Figure \ref{fig:espectro} (a) with for the analytical description (black line) and with for the experimental results (red line).

In the deep BSA case, Polyethylene was employed as neutron moderator material as well. As low energy filter different chlorides were tested, we chose Hexachlorobenzene (C$_6$Cl$_6$) to scatter low energy neutrons, where the Carbon and Chlorine scattering cross-sections are much lower on high neutron energies. In addition, Hexachlorobenzene compound is a solid extensively used on the agrarian industry, thus it could be easy its manipulated and acquired. To minimise the fast neutrons, we used Titanium (Ti); and Lead (Pb) was employed as neutron reflector material. Using MCNPX \cite{mcnpx} to perform the design, a combination of these four materials was optimised for the treatment of deep tumours with $^7$Li(p,n) reaction at $1950~$keV. The main dimensions of the BSA are shown in Table \ref{tab:bsa_design}. The patient was placed at $1~$cm to the BSA. With a proton current of $85~$mA over a Lithium-metallic sample, this work is in agreement with the IAEA neutron figures of merit recommendations for deep tumours. The neutron flux-energy spectrum in the tumour area is shown in Figure \ref{fig:espectro} (b), where a black line represents the result obtained by means of the analytical description and a red line the result obtained via experimental results of the neutron field.

\begin{table}[ht]
\centering
\footnotesize
 \begin{tabular}{c|cccc}
   \toprule
BSA & Element & Material & Length (cm) & Radio (cm)\\
  \midrule
  
\multirow{2}{*}{\rotatebox[origin=c]{90}{Shallow}} & Moderator & Polyethylene & 23.4 & 10 \\
& Reflector & Lead & 25.4 & 40 \\

\midrule

\multirow{4}{*}{\rotatebox[origin=c]{90}{Deep}} & Moderator & Polyethylene & 8.9 & 10 \\
& Low Energy Filter & C$_6$Cl$_6$ & 5.1 & 10 \\
& High Energy Filter & Ti & 4 & 10 \\
& Reflector & Lead & 19 & 40 \\
\bottomrule
 \end{tabular}
 \caption{Materials and dimensions of the BSA for shallow and deep tumours. Each element presents a cylinder shape.}
 \label{tab:bsa_design} 
\end{table}

\begin{table}[ht]
\centering
\footnotesize
 \begin{tabular}{c|cccc}

\toprule
\multirow{2}{*}{BSA} & IAEA & IAEA & this work & this work \\
& figures of merit & thresholds & (L\&Z+BSA) & (Measure+BSA) \\
\midrule 
 
\multirow{3}{*}{\rotatebox[origin=c]{90}{\begin{tabular}{c}Shallow\\($81~$mA)\\ \end{tabular}}} & $\Phi_{th}~$(cm$^{-2}$s$^{-1}$) & $\geqslant 10^9$ & $10^9$ & $10^9$ \\
& $\Phi_{th}/\Phi_{tot}$ & $\geqslant 0.9$ & 0.932 & 0.933 \\
& \.{D}$_{epi+fast}/\Phi_{th}~$(Gy\,cm$^2$) & $\leqslant 2 \times 10^{-13}$ & $1.99 \cdot 10^{-13}$ & $1.93 \cdot 10^{-13}$ \\
\midrule 

\multirow{4}{*}{\rotatebox[origin=c]{90}{\begin{tabular}{c}Deep\\($85~$mA)\\ \end{tabular}}} & $\Phi_{th}~$(cm$^{-2}$s$^{-1}$) & $\geqslant 10^9$ & $10^9$ & $10^9$ \\
& $\Phi_{epi}/\Phi_{th}$ & $\geqslant 100$ & 100.51 & 102.27 \\
& $\Phi_{epi}/\Phi_{fast}$ & $\geqslant 20$ & 20.01 & 20.60 \\
& \.{D}$_{fast}/\Phi_{epi}~$(Gy\,cm$^2$) & $\leqslant 2 \times 10^{-13}$ & $ 1.63 \cdot 10^{-13}$ & $1.68 \cdot 10^{-13}$ \\
\bottomrule
 \end{tabular}
 \caption{Comparison of the IAEA neutron figures of merit for shallow and deep tumours and this work with $81~$mA and $85~$mA proton currents respectively. This work is composed by the analytical description of the neutron field, proposed by Lee \& Zhou (L\&Z), and the experimental results of the measure carried out at HiSPANoS facility (Measure).}
 \label{tab:bsa_IAEA} 
\end{table}

\section{Conclusions}
In this study, two BSAs for AB-BNCT have been designed and optimized for a neutron source based on the $^7$Li(p,n) reaction at $1950~$keV. We have employed the neutron field provided by the experimental results of the TOF measurement performed at HiSPANoS facility, in agreement with the Lee \& Zhou \cite{Lee} analytical description. The BSAs, proposed in this paper, suggest a therapeutic result for BNCT of shallow and deep tumours when compared to the IAEA neutron figures of merit, with a lower proton energy load to the Li target than previously reported and reasonable proton currents.

\section{Acknowledgments}
The authors thank CNA technicians for their help during the experiment. We acknowledge the financial support for this work from the Spanish MINECO and FEDER funds under contract FIS2015-69941-C2-1-P, FPA2016-77689-C2-1-R, A-FQM-371-UGR18 (Programa Operativo FEDER Andaluc\'ia 2014–2020), Asociaci\'on Espa\~nola Contra el C\'ancer (AECC) under contract PS16163811PORR, Junta de Andaluc\'ia under contract P11-FQM-8229, and the founders of the University of Granada Chair Neutrons for Medicine: Spanish Fundaci\'on ACS and Capit\'an Antonio. M. Mac\'ias acknowledges the FPI fellowship BES-2014-068808 granted in the project FPA2013-47327-C2-1-R.


\begin{thebibliography}{10}
\expandafter\ifx\csname url\endcsname\relax
  \def\url#1{\texttt{#1}}\fi
\expandafter\ifx\csname urlprefix\endcsname\relax\def\urlprefix{URL }\fi
\expandafter\ifx\csname href\endcsname\relax
  \def\href#1#2{#2} \def\path#1{#1}\fi

\bibitem{bnctref2}
R.~{Barth {\sl et al.}}, \href{https://doi.org/10.1186/1748-717X-7-146}{Current
  status of boron neutron capture therapy of high grade gliomas and recurrent
  head and neck cancer}, Radiation Oncology 7~(1) (2012) 146.
\newblock \href {http://dx.doi.org/10.1186/1748-717X-7-146}
  {\path{doi:10.1186/1748-717X-7-146}}.
\newline\urlprefix\url{https://doi.org/10.1186/1748-717X-7-146}

\bibitem{ab_ref5}
S.~{Halfon {\sl et al.}},
  \href{http://www.sciencedirect.com/science/article/pii/S096980431100159X}{High-power
  liquid-lithium target prototype for accelerator-based boron neutron capture
  therapy}, Applied Radiation and Isotopes 69~(12) (2011) 1654--1656, special
  Issue: 14th International Conference on Neutron Capture Therapy.
\newblock \href
  {http://dx.doi.org/https://doi.org/10.1016/j.apradiso.2011.03.016}
  {\path{doi:https://doi.org/10.1016/j.apradiso.2011.03.016}}.
\newline\urlprefix\url{http://www.sciencedirect.com/science/article/pii/S096980431100159X}

\bibitem{ABNS-BNCT1}
Y.~{Oka {\sl et al.}}, \href{https://doi.org/10.13182/NT81-A32809}{A design
  study of the neutron irradiation facility for boron neutron capture therapy},
  Nuclear Technology 55~(3) (1981) 642--655.
\newblock \href {http://arxiv.org/abs/https://doi.org/10.13182/NT81-A32809}
  {\path{arXiv:https://doi.org/10.13182/NT81-A32809}}.
\newline\urlprefix\url{https://doi.org/10.13182/NT81-A32809}

\bibitem{ABNS-BNCT2}
J.~{Blue {\sl et al.}}, A study of low energy proton accelerators for neutron
  capture therapy., Proceedings of Second International Symposium on Neutron
  Capture Therapy. (1985) 147.

\bibitem{ABNS-BNCT3}
T.~{Wangler {\sl et al.}}, Conceptual design of an rfq accelerator-based
  neutron source for boron neutron-capture therapy, Proceedings of the 1989
  IEEE Particle Accelerator Conference, . 'Accelerator Science and Technology
  (1989) 678--680 vol.1.

\bibitem{nt}
Neutron Therapeutics, Inc., \url{https://www.neutrontherapeutics.com/} (2020).

\bibitem{shi}
Sumitomo Heavy Industries, Ltd., \url{https://www.shi.co.jp/english/index.html} (2020).

\bibitem{ray}
RaySearch Laboratories, \url{https://www.raysearchlabs.com/} (2020).

\bibitem{bnctref1}
A.~{Kreiner {\sl et al.}},
  \href{http://www.sciencedirect.com/science/article/pii/S1507136714001837}{Present
  status of accelerator-based bnct}, Reports of Practical Oncology and
  Radiotherapy 21~(2) (2016) 95--101, 7th Young BNCT meeting.
\newblock \href {http://dx.doi.org/https://doi.org/10.1016/j.rpor.2014.11.004}
  {\path{doi:https://doi.org/10.1016/j.rpor.2014.11.004}}.
\newline\urlprefix\url{http://www.sciencedirect.com/science/article/pii/S1507136714001837}

\bibitem{bnctnicola0}
E.~Bisceglie, P.~Colangelo, N.~Colonna, P.~Santorelli, V.~Variale,
  \href{https://doi.org/10.1088%2F0031-9155%2F45%2F1%2F304}{On the optimal
  energy of epithermal neutron beams for {BNCT}}, Physics in Medicine and
  Biology 45~(1) (1999) 49--58.
\newblock \href {http://dx.doi.org/10.1088/0031-9155/45/1/304}
  {\path{doi:10.1088/0031-9155/45/1/304}}.
\newline\urlprefix\url{https://doi.org/10.1088%2F0031-9155%2F45%2F1%2F304}

\bibitem{bnctnicola}
E.~{Bisceglie {\sl et al.}},
  \href{http://www.sciencedirect.com/science/article/pii/S0168900201014061}{Production
  of epithermal neutron beams for bnct}, Nuclear Instruments and Methods in
  Physics Research A 476~(1) (2002) 123--126, int. Workshop on Neutron Field
  Spectrometry in Science, Technolog y and Radiation Protection.
\newblock \href
  {http://dx.doi.org/https://doi.org/10.1016/S0168-9002(01)01406-1}
  {\path{doi:https://doi.org/10.1016/S0168-9002(01)01406-1}}.
\newline\urlprefix\url{http://www.sciencedirect.com/science/article/pii/S0168900201014061}

\bibitem{Lee}
C.~Lee, X.-L. Zhou, Nuclear Instruments and Methods in Physics Research B 152
  (1999) 1--11.

\bibitem{LiskienP}
H.~Liskien, A.~Paulsen, Atomic Data and Nuclear Data Tables 15 (1975) 57--84.

\bibitem{mcnpx}
D.~Pelowitz, MCNPX Users Manual Version 2.5.0, Los Alamos National Laboratory
  LACP, 2005.

\bibitem{CNA}
J.~{Garc\'ia-L\'opez {\sl et al.}}, Cna: The first accelerator-based iba
  facility in spain, Nuclear Instruments and Methods in Physics Research B
  161-163 (2000) 1137--1142.
\newblock \href {http://dx.doi.org/10.1016/S0168-583X(99)00702-8}
  {\path{doi:10.1016/S0168-583X(99)00702-8}}.

\bibitem{macias0}
M.~Mac\'ias, B.~Fern\'andez, J.~A. Labrador, A.~Romero, J.~Praena, Il Nuovo
  Cimento 42 C (2019) 63.

\bibitem{MACIAS}
M.~Mac\'ias, B.~Fern\'andez, J.~Praena,
  \href{http://www.sciencedirect.com/science/article/pii/S0969806X19302968}{The
  first neutron time-of-flight line in spain: Commissioning and new data for
  the definition of a neutron standard field}, Radiation Physics and Chemistry
  168 (2020) 108538.
\newblock \href
  {http://dx.doi.org/https://doi.org/10.1016/j.radphyschem.2019.108538}
  {\path{doi:https://doi.org/10.1016/j.radphyschem.2019.108538}}.
\newline\urlprefix\url{http://www.sciencedirect.com/science/article/pii/S0969806X19302968}

\bibitem{dt}
CAEN~SpA, User Manual UM3148 DT5730 / DT5725, https://www.caen.it/, 2016.

\bibitem{coinci}
CAEN~SpA, User Guide GD2827 How to make coincidences with CAEN Digitizers,
  https://www.caen.it/, 2017.

\bibitem{liglass}
Scionix Ltd., \url{https://scionix.nl/} (2019).



\end{thebibliography}
\end{document}